\documentclass[11pt]{iopart}
\pdfoutput=1
\usepackage{graphics}
\usepackage{tabularx}
\usepackage{mathrsfs}
\usepackage[dvips]{graphicx} 
\usepackage{amssymb}
\expandafter\let\csname equation*\endcsname\relax
\expandafter\let\csname endequation*\endcsname\relax
\usepackage{amsmath}
\usepackage{epsfig}
\usepackage{color}
\usepackage{ifthen}
\usepackage{multirow} 	
\usepackage[usenames,dvipsnames]{xcolor}
\usepackage{pdfpages}
\usepackage[colorlinks=true,urlcolor=blue,linkcolor=blue,citecolor=blue,linktocpage=true]{hyperref}
\usepackage{varioref}
\makeatletter
\renewcommand\@dotsep{200}
\makeatother
\begin{document}
\title[Tensor Minkowski Functionals: first application to the CMB]{Tensor Minkowski Functionals: first application to the CMB}

\author{Vidhya Ganesan$^{1,2}$} 
\ead{vidhya@iiap.res.in} 

\author{Pravabati Chingangbam$^{1}$} 
\ead{prava@iiap.res.in} 

\address{$^1$ Indian Institute of Astrophysics, Koramangala II Block, Bangalore  560 034, India\\
  $^2$Indian Institute of Science, C V Raman Ave, Bangalore 560 012, India 
} 
\begin{abstract}
Tensor Minkowski Functionals (TMFs) are tensorial generalizations of the usual Minkowski Functionals which are scalar quantities.  We introduce them here for use in cosmological analysis, in particular to analyze the Cosmic Microwave Background (CMB) radiation. They encapsulate information about the shapes of structures and the orientation of distributions of structures. We focus on one of the TMFs, namely $W_2^{1,1}$, which is the $(1,1)$ rank tensor generalization of the genus.  The ratio of the eigenvalues of the average of $W_2^{1,1}$ over all structures, $\alpha$, encodes the net orientation of the structures; and the average of the ratios of the eigenvalues of $W_2^{1,1}$ for each structure, $\beta$, encodes the net intrinsic anisotropy of the structures. We have developed a code that computes $W_2^{1,1}$, and from it $\alpha$ and $\beta$, for a set of structures on the 2-dimensional Euclidean plane. We use it to compute $\alpha$ and $\beta$ as functions of chosen threshold levels for simulated Gaussian and isotropic CMB temperature and $E$ mode fields. We obtain the value of $\alpha$ to be one for both temperature and $E$ mode, which means that we recover the statistical isotropy of density fluctuations that we input in the simulations. We find that the standard inflationary $\Lambda$CDM predicts that the level of intrinsic  anisotropy of hotspot and coldspot structures in the CMB fields is quantified by $\beta\sim 0.62$. Then we compute $\alpha$ and $\beta$ for temperature and $E$ mode data from the PLANCK mission. We find that the temperature field agrees with the standard $\Lambda$CDM prediction of no net orientation within $3-\sigma$. However, we find that the structures in $E$ mode data have a net orientation that deviates from the theoretical expectation at $14-\sigma$. The possible origin of this deviation may be due to instrumental effects or other sources and needs to be investigated further. For the net intrinsic anisotropy of structures we obtain values of $\beta$ for both temperature and $E$ mode that are consistent with the expectations from the standard $\Lambda$CDM simulations. Accurate measurements of $\alpha$ and $\beta$ can be used to test the standard model of cosmology and to search for deviations from it.
\end{abstract}

\maketitle
\section{Introduction}
From the time of its first detection by Penzias and Wilson in 1965 \cite{penzias:1965}, the cosmic microwave background (CMB) radiation   has provided us with a wealth of information about the Universe. The information is encoded in the statistical properties of temperature and polarization anisotropies~\cite{Bond:1987ub}. Fluctuations in the CMB temperature were first detected by COBE~\cite{smoot:1992}.  
Less than 10$\%$ of CMB photons are linearly polarized due to the presence of quadrupole anisotropies in the plasma just before the matter-radiation decoupling phase \cite{Coulson:1994}. This polarization was first detected by DASI \cite{Kovac:2002fg}.  
Polarization observations measure the Stokes parameters $Q$ and $U$. They are however not invariant under rotations about the lines of sight. The complex quantities $Q\pm iU$ transform as spin $\mp 2$ objects, respectively. These are usually expressed in terms of gradient component and curl component known as $E$ mode and $B$ mode respectively. As $E$ mode and $B$ mode transform as scalars under these rotation transformations, these fields are widely used for the polarization analysis. The PLANCK team has recently released high resolution map of $E$ mode \cite{adam:2015}. The additional information contained in the polarization data has significantly improved the constraints on cosmological parameters. The main contribution of $E$ mode polarization comes from the primordial density fluctuations \cite{Bardeen:1983} while the $B$ mode polarization can be generated by primordial gravitational waves \cite{Starobinsky:1979} or secondary effects such as gravitational lensing of CMB photons by large scale structure. $B$ mode polarization of primordial origin has not been detected yet, while the component that is sourced by lensing has been detected by SPTpol \cite{hanson:2013}.  

CMB fluctuations have been analyzed so far using various tools such as power spectrum, bispectrum, trispectrum, scalar Minkowski Functionals (SMFs) and so on. SMFs are used to study the geometric and topological properties of cosmological fields \cite{Tomita:1986,gott:1990,mecke:1994,sch:1997,winit:1998,mat:2003,hik:2006,mat:2010,Schmalzing:1998}. It has been applied to study primordial non-Gaussianity using CMB temperature \cite{wmap:2008,planck:2013} and $E$ mode fields \cite{planck:2015,Ganesan:2014lqa}. To name a few other applications, SMFs have also been used to study the effect of foreground contamination \cite{fore:2013,bmode:2015}, cold spot anomaly \cite{cold:2014} and modified gravity theories through lensing \cite{lens:2016} in CMB. Vector extensions of SMFs have been introduced into cosmology and applied to study the morphology and dynamical evolution of galaxy clusters \cite{beis:2000,beis:2001}. Some of the other methods that have been used to study the topological properties are through clustering strength \cite{ross:2009,ross:2010,ross:2011}, number of hot and cold spots \cite{coles:1987,prava:2012,park:2013,Ganesan:2014lqa} and extrema counts \cite{pogo:2009}. 

The observed CMB data is consistent with the standard model of cosmology. However, there are a few anomalies seen in the CMB temperature anisotropy that contradict the expectations from the standard model. Some of these anomalies are low variance, hemispherical asymmetry, point parity asymmetry, mirror parity asymmetry and cold spot \cite{ade:2015}.  These have been detected by both WMAP and PLANCK experiments, implying that their origin is not due to the systematics in these experiments. The PLANCK team's final data release which will include the analysis of large angular scale polarization will further reveal the statistical significance of these anomalies. In order to get a better understanding and resolution of these anomalies it is desirable to develop efficient methods which can analyze the data. 

Tensor Minkowski Functionals (TMFs) are $(a,b)$ rank tensorial extensions of SMFs \cite{Alesker:1999,Hug:2008} that are constructed by taking tensor product of $a$ copies of the position vector and $b$ copies of the unit normal at each point of the boundary of a given structure (see section 2 for detailed mathematical definition). By virtue of being tensor quantities and having more degrees of freedom, they carry additional information in comparison to SMFs about the morphology of a single structure or a set of many structures. In this paper we introduce TMFs as tools to analyze cosmological fields, in particular for analyzing the CMB fields which are defined on two dimensions. In this work we focus attention on one of the TMFs, denoted by $W_2^{1,1}$, which is a generalization of the genus (one of the three SMFs in two dimensions), and which is a $(1,1)$ rank tensor. For a given set of many structures there are two ways of obtaining the eigenvalues of $W_2^{1,1}$. The first is to average the elements of $W_2^{1,1}$ over all the structures and then calculate the eigenvalues. The ratio of the two eigenvalues gives a measure of the degree of alignment or relative orientation of the structures. The second way is to calculate the eigenvalues for each structure, take their ratio, and then obtain the average ratio over all the structures. The average of the ratio of eigenvalues calculated in the second way gives a measure of the average intrinsic anisotropy of the structures. These new information about net orientation and net intrinsic anisotropy of a set of structures can be useful in resolving the anomalies in the CMB data, test the standard model of cosmology and search for deviations from it, and to discriminate different early Universe physics models. They can also be used to study characteristic signatures of instrumental and foreground signals. Moreover, TMFs are quite promising for analyzing data of the large scale structure of the Universe and 21 cm emissions from the epoch of reionization and extracting cosmological information from them. 

G. E. Schroder-Turk {\em et. al.} \cite{schroder2D:2009} have provided with an explicit formulae to calculate TMFs for structures on two dimensions. We implement the formulae and develop a code, that we refer to as TMFCode, to calculate TMFs, in particular $W_2^{1,1}$, and from it the two kinds of ratio of eigenvalues described above. We carry out a detailed analysis of the numerical inaccuracies in these quantities due to pixelization of the plane. For application to the CMB fields we use the stereographic projection to project the CMB data which is given on the spherical sky, onto a Euclidean plane to calculate the TMFs. We first calculate $W_2^{1,1}$ and the two eigenvalue ratios for simulated CMB fields in order to calculate the net orientation and net anisotropy of structures predicted by the standard $\Lambda$CDM cosmology. Then we apply to temperature and $E$ mode polarization data from PLANCK. We find no net orientation in the temperature data in agreement with the standard $\Lambda$CDM prediction of no net orientation within $3-\sigma$. However, we find $14-\sigma$ evidence for a net orientation in $E$ mode. This deviation may be due to instrumental effects or other sources and requires  further investigation for a proper understanding. We will pursue it further after the full PLANCK data release.   
For the net intrinsic anisotropy of structures we obtain values for both temperature and $E$ mode that are consistent with the expectations from the standard model. 

TMFs have been used in various research areas to study a wide range of phenomena ranging from the shape of neuronal cells in the brain \cite{Beisbart:2006}, to the shape of ice crystals in Antarctica \cite{Durand:2004}, to the shape of galaxies and clusters of galaxies in our Universe \cite{Beisbart:2002}. 

This paper is organized as follows. In section \ref{sec2} we give the definition of tensor Minkowski Functionals, describe its numerical calculation and the corresponding numerical inaccuracies due to the pixelization of the data. In section \ref{sec3} we apply the TMF $(W_2^{1,1})$ to simulated CMB fields and we calculate the variation of net orientation $(\alpha)$ and net anisotropy $(\beta)$ with the threshold value, and we also give an estimate of $\alpha$ and $\beta$ that we expect for a Gaussian, isotropic CMB field. In section \ref{sec4} we apply the TMF $(W_2^{1,1})$ to PLANCK data, we compare the values of $\alpha$ and $\beta$ with that of the simulated CMB fields and estimate the level of consistency of PLANCK data with the standard model of cosmology. In appendix A, we describe how to quantify the level of orientation for a set of structures on the sphere.  Finally, we summarize the results along with a discussion of their implications and future directions in section \ref{sec5}.

\section{Tensor Minkowski Functionals} \label{sec2}
In this section we give the mathematical definition of tensor Minkowski Functionals and describe their numerical calculation for 2-dimensional random fields on Euclidean space.

\subsection{Definition}
Let us denote a subset of 2-dimensional Euclidean space with smooth closed boundary (or boundaries) by $K$, and let the boundary contour (or contours) be denoted by $\partial K$. $K$ may consist of one or more simply and/or multiply connected regions. A simply connected region will have no hole and a multiply  connected region may have one or a countable number of holes in them. To connect with the terminology used in geometrical and topological analyses of CMB fields, which is our main interest here, we refer to $K$ as an {\em excursion set}. A connected region is referred to as a {\em hotspot} and a hole as a {\em coldspot}. 
We will use the word {\em structure} to mean either a connected region or a hole. The contour that encloses a hotspot is assigned anticlockwise direction while a contour that encloses a hole is assigned clockwise direction. 

The morphological properties of $K$ can be characterized by three scalar Minkowski Functionals (SMFs) which are defined as
\begin{equation} 
W_0(K)= \int_K d^2r, \quad  W_{i}(K)=\frac{1}{2} \int_{\partial K} G_i dr,
\label{eqn:smf}
\end{equation}
where $i = 1,2$, and the functions $G_i$ are $G_1=1$ and $G_2=\kappa$ \cite{gott:1990, Schmalzing:1998}. The symbol $\kappa$ denotes the local curvature at each point on $\partial K$. Physically $W_0, W_{1}$ are the area and the length of the boundary of connected region of $K$ and $W_2$ is the number of connected regions minus the number of holes in $K$.  
Note that these definitions have normalization constants that are different from the usual SMFs used in the CMB analysis. However, their physical interpretation remains the same. Being scalar quantities they are invariant under coordinate transformations and insensitive to the orientation and anisotropy of the structures (connected regions and holes). 

Tensor Minkowski Functionals are defined by generalizing the SMFs~\cite{Alesker:1999,Hug:2008}, as follows. For simplicity let us focus on a single simply or multiply connected region $K_s \subset K $. The tensor Minkowski Functionals of rank $a+b$, with $a+b\ge 0$, are constructed by taking $a$ number of copies of the position vector $\vec{r}$ and $b$ number of copies of the normal vector $\vec{n}$ at each point on the contour $\partial K_s$ and taking their tensor product. The tensor Minkowski Functionals of $K_s$ with rank $a+b$ are then defined as
\begin{equation} 
  \begin{array}{l}
   \displaystyle W_0^{a,0}(K_s) =  \int_{K_{s}} \vec{r}^{\,a} d^2 r\\
    \displaystyle W_{\nu}^{a,b}(K_s) = \frac{1}{2} \int_{\partial K_{s}} \vec{r}^{\,a} \otimes \vec{n}^{\,b} G_i dr,
  \end{array}
  \label{eqn:tmf}
\end{equation}
for $i = 1,2$. The functions $G_i$ are the same as in Eq.~(\ref{eqn:smf}). The tensor product between two vectors $\vec{A}$ and $\vec{B}$ is given by $(\vec{A} \otimes \vec{B})_{ij}=(A_iB_j+A_jB_i)/2$. For $a+b=0$, Eq.~(\ref{eqn:tmf}) reduces to  the three SMFs. For $a=1$ and $b=0$ we get three vectorial Minkowski Functionals, and for $a+b=2$ we get seven tensor Minkowski Functionals of rank 2. 

In general, the tensor Minkowski Functionals, unlike their scalar counterparts, are not invariant under coordinate transformations. They behave like tensors of corresponding rank. Specifically, tensors $W_1^{1,1}, W_1^{0,2}, W_2^{1,1}$ and $W_2^{0,2}$ are invariant under translation operation while others vary. 
In this paper we focus only on translation invariant tensors as translation covariant tensors are sensitive to position or the choice of origin. Specifically, we choose the tensor $W_{2}^{1,1}$ for the study as other translation invariant tensors are related to either $W_2^{1,1}$ or $W_{\nu}$.

In practice real data is pixelized and hence in this case the excursion set $K$ will be a finite set consisting of pixels that are included in the structures. $K_s \subset K$ will be a finite set of pixels which form a single structure. Therefore, the boundary of a single simply  connected region will be a polygon, a doubly connected region (having one hole) will consist of two polygons, with the one corresponding to the hole being located inside the external one, and so on.     
The formulae for calculating TMFs of each polygon of $K_s$ is given  in~\cite{schroder2D:2009}. In particular, the tensor $W_2^{1,1}$ is given by
\begin{equation}
\begin{array}{l}
  \displaystyle W_2^{1,1}(K_s) = \sum_{(i,j)} \frac{1}{2} |e_{ij}|^{-1} (\vec{e}_{\,ij} \otimes \vec{e}_{\,ij}),\\
\end{array}
\label{eq:for}
\end{equation}
where the pair $(i,j)$ labels the edge of the polygon between $i$ and $j$ vertices. $\vec{e}_{\,ij}$ is the edge of the polygon with length $|\vec{e}_{\,ij}|$. As mentioned above, the direction of $\vec{e}_{\,ij}$ is such that the boundary of a hotspot  is anticlockwise while the boundary of a coldspot is clockwise. 

\subsection{Measure of anisotropy and orientation of structures}

For each structure (hotspot or coldspot) of $K$, $W_2^{1,1}$ being a $2\times 2$ matrix has two real eigenvalues, which we denote by $\lambda_1$ and $\lambda_2$ such that $\lambda_1 \le \lambda_2$. For the entire $K$ let $\langle \dots \rangle$ denote averaging over all structures (hotspots or coldspots). Then let $\Lambda_1$ and $\Lambda_2$, such that $\Lambda_1\le \Lambda_2$, denote the eigenvalues of  $\left\langle  W_2^{1,1} \right\rangle$, where the averaging is done for each element of $W_2^{1,1}$. 
Then, we define the ratios $\alpha$ and $\beta$ as  
\begin{equation}
\alpha\equiv   \frac{\Lambda_1}{\Lambda_2}, \quad 
\beta\equiv\left\langle  \frac{\lambda_1}{\lambda_2}\right\rangle.
\label{eq:avg}
\end{equation}
As observed from the above formula, $\alpha$ and $\beta$ are both ratios of eigenvalues but they differ in the averaging process. $\alpha$ is obtained by first averaging $W_2^{1,1}$ over all structures, calculating the eigenvalues of $\left\langle  W_2^{1,1} \right\rangle$ and then calculating their ratio. On the other hand $\beta$ is obtained by first calculating the ratio of eigenvalues of $W_2^{1,1}$ for each structure and then averaging the ratio over all the structures.

If the excursion set consists of only one structure then $\alpha=\beta$.
For a given structure $\beta$ gives a measure of {\em the intrinsic anisotropy} or the deviation from rotational symmetry in the shape of the structure. For the case of many structures it gives the net anisotropy of all the structures. It is a straightforward exercise to show that $\beta=1$ for isotropic shapes such an equilateral triangle, square and circle. For a general shape the value of $\beta$ lies between 0 and 1, 
with $\beta<1$ implying that the shape of the structure is anisotropic with the value indicating the extent of anisotropy in the shape of the structure. 

The arrangement of many anisotropic structures can be associated with an {\em orientation}. $\alpha$ gives a measure of the orientation or the deviation from rotational symmetry in the distribution of structures. If the structures are oriented randomly with no preferred direction then $\alpha=1$, else $\alpha$ lies between 0 and 1.  
If the structures with certain fixed anisotopic shape are arranged such that they are all orientated along the same direction then $\alpha$ will be equal to $\beta$.
We define the quantity $\mathscr{O}$ which is a measure of the degree of alignment of the structures as
\begin{equation}
\mathscr{O}\equiv\frac{1-\alpha}{1-\beta}.
\label{eqn:O}
\end{equation}
$\mathscr{O}=1$ means that the structures are completely aligned, while  $\mathscr{O}=0$  means no net orientation. Any value between zero and one will indicate the extent to which the structures are orientated. As this is a normalized form of net orientation $\alpha$, this will be a useful quantity for the purpose of comparison of the extent of orientation in the structures between different fields. 

\subsection{Numerical calculation of TMFs}
We have developed a code that computes TMFs of individual structures of an  excursion set on a plane and then estimates the ratio of eigenvalues $\alpha$, $\beta$ of these structures. We refer to this code as TMFCode. When an excursion set is input to the TMFCode it first estimates the line segments which forms the edges of the structures at each vertex (point where four neighbouring pixels meet) on the pixelized plane based on the configurations of the surrounding pixels (see Fig.~7 of \cite{schroder2D:2009}). The details of each structure such as $\vec{e}_{\,ij}$ corresponding to each of its edges are stored. Then the code tracks all of the individual structures in the excursion set and assigns labels to each of these structures which allows us to retrieve the details of an individual structure whenever necessary.  
Then $W_2^{1,1}$ can be calculated for each of these structures using Eq.~\ref{eq:for}, and from it $\alpha$ and $\beta$ are calculated.

\subsubsection{Anisotropy of a single elliptical disk on a plane and estimation of pixelization error:}

\begin{figure}
\begin{center}
\resizebox{2in}{2in}{\includegraphics{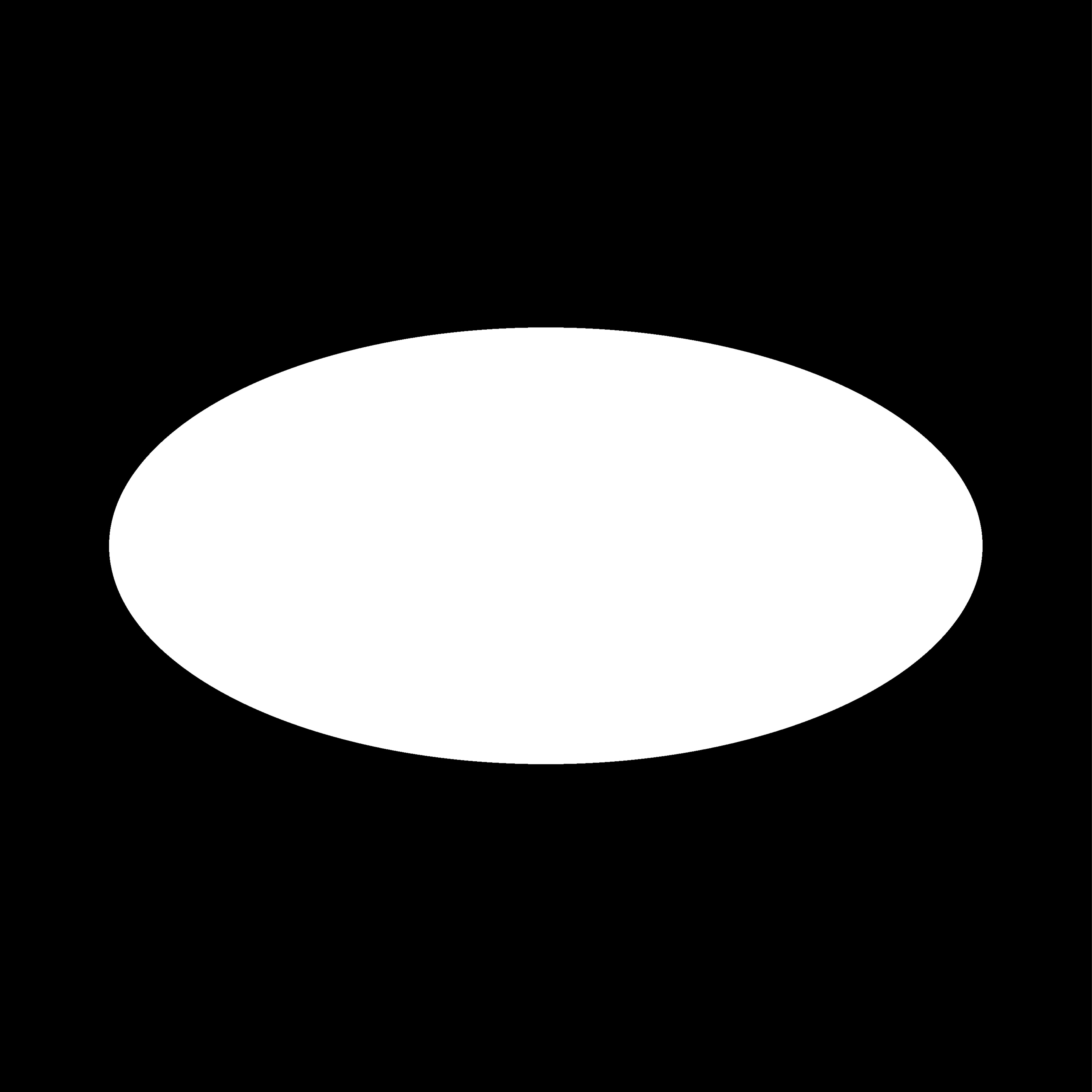}}\quad
\end{center}
\caption{The white region shows an excursion set consisting of a single elliptical disk.}
\label{fig:ellipse}
\end{figure}

The approximation of a continuous contour by a polygon on a pixelized space 
will inherently lead to numerical inaccuracy in the calculation of $W_2^{1,1}$. 
We quantify this for an excursion set consisting of a single elliptic disk (see Fig.~\ref{fig:ellipse}) for which $W_2^{1,1}$ is given by the following analytic expression,
\begin{equation}
\begin{array}{l}
\displaystyle W_2^{1,1}= \begin{bmatrix} f_2^{1,1}(p,q) & 0 \\ 0 & f_2^{1,1}(q,p), \end{bmatrix} \\
\vspace{0.2cm} \\
\displaystyle f_2^{1,1}(p,q)=\frac{1}{2} p^2 q^2 \int_0^{2\pi} d \varphi \frac{cos^2 \varphi}{\left[ p^2-(p^2-q^2)cos^2 \varphi \right]^{3/2}},
\end{array}
\label{eq.tensorformula}
\end{equation}
where $p$ and $q$ are the semi-major and semi-minor axes, respectively.

\begin{table}
  \centering{}
  \begin{tabular}{| p{1.7cm} | p{1.7cm} | c | c | c | p{1.7cm} |}
    \hline
    $q/p$  &  $\beta$ from & \multicolumn{3}{c|}{$\beta$ from TMFCode} & $\%$ error \\
    & analytical & \multicolumn{3}{c|}{} & \\
    & formula & \multicolumn{3}{c|}{} & \\
   \cline{3-5}
 & & $1000^2$ & $2000^2$ & $3000^2$ &\\
& & pixels & pixels & pixels & \\
\hline
1.0000 & 1.0000 & 1.0000 & 1.0000 & 1.0000 & 0.0 \\
\hline 
0.8000 & 0.7154 & 0.7642 & 0.7641 & 0.7641 & 6.8 \\
\hline
0.6000 & 0.4638 & 0.5418 & 0.5417 & 0.5418 & 16.8 \\
\hline 
0.5000 & 0.3518 & 0.4371 & 0.4370 & 0.4370 & 24.2 \\
\hline
0.3000 & 0.1602 & 0.2432 & 0.2433 & 0.2432 & 51.8 \\
\hline 
0.1000 & 0.0274 & 0.0739 & 0.0741 & 0.0741 & 170.4 \\
\hline
  \end{tabular}
  \caption{Values of $\beta$ for a single elliptical disk, with different values of the aspect ratio $q/p$ (column 1), obtained from the analytical formula Eq.~\ref{eq.tensorformula} (column 2) and from the numerical calculation using TMFCode (column 3). Column 4 shows the percentage numerical error in $\beta$ corresponding to $3000^2$ pixels. }
\label{table:plane}
\end{table}
Table~\ref{table:plane} summarizes the values of $\beta$ for a continuous ellipse obtained using Eq.~\ref{eq.tensorformula} and the corresponding values for a polygon obtained from the TMFCode that uses Eq.~\ref{eq:for}, for different resolutions corresponding to $1000^2,\ 2000^2$ and $3000^2$ pixels. Note that $\beta=\alpha$ in this case since there is only one structure. 
For a fixed aspect ratio, we find that the value of $\beta$ does not vary significantly with the resolution.
However for a fixed resolution, we find that the numerical inaccuracy in $\beta$ increases as the aspect ratio $q/p$ decreases. This is because the polygon edges fails to capture the true curvature at 
every point of the contour of the continuous ellipse. Hence the cause of this numerical inaccuracy is the pixelization of the continuous ellipse. The parts of the contour having higher curvature will lead to larger error. 
We can notice that the polygons systematically over-estimate the value of $\beta$ in comparison to the analytically expected value of $\beta$ for the continuous ellipse. We will use this fact to make approximate corrections due to the pixel effect when we apply to the CMB fields.  

\subsubsection{Orientation of two elliptical disks on a plane and estimation of pixelization error:}

In order to quantify the numerical error in the calculation of $\alpha$ due to the pixelization we compare $\alpha$ obtained for the excursion set consisting of two continuous ellipses and the corresponding value obtained with its approximation as two polygons on a pixelized space. We fix the aspect ratio $q/p$ of both the ellipses to be 0.5.
Table. \ref{tle:orient} shows $\alpha$ for two continuous ellipses obtained using Eq.~\ref{eq.tensorformula}, and the corresponding values for two polygons obtained from the TMFCode which uses Eq.~\ref{eq:for}, for various relative orientations of the major axes of the two ellipses. $0^{\circ}$ corresponds to the case where the ellipses are completely aligned with each other, while $90^{\circ}$ corresponds to the case where the major axes are perpendicular to each other, and hence are completely unoriented with each other. For a fixed relative orientation, we find that the value of $\alpha$ does not vary significantly with the resolutions. However, for a fixed resolution, we find that the numerical inaccuracy decreases as the two ellipses become more and more unoriented with each other. When both the ellipses are completely aligned then $\alpha = \beta$. Hence we find that the numerical inaccuracy in $\alpha$ for the case where the ellipses are completely oriented is the same as the corresponding numerical inaccuracy in $\beta$. But as the ellipses become more and more unoriented the value of $\alpha$ approaches one irrespective of the value of $\beta$, hence the numerical inaccuracy also decreases. Therefore, for the application to real data with $\alpha$ close to one, the corrections to $\alpha$ due to the pixel effect can be neglected as the numerical inaccuracy in $\alpha$ is not significant when $\alpha$ is close to one. For the case when the real data has $\alpha$ close to $\beta$, then the corrections due to the pixel effect has to be taken into account. We will use this result when we apply to the CMB fields.

\begin{table}
  \centering{}
  \begin{tabular}{|c|c|c|c|c|c|}
    \hline
    Angle between & $\alpha$ from & \multicolumn{3}{c|}{$\alpha$ from TMFCode} & $\%$ error \\
    major axis & analytical & \multicolumn{3}{c|}{} & \\
    of the ellipses & formula & \multicolumn{3}{c|}{} & \\
    \cline{3-5}
    && $1000^2$ & $2000^2$ & $3000^2$ & \\
    && pixels & pixels & pixels & \\
    \hline
    $0^{\circ}$ & $0.3518$ & $0.4369$ & $0.4371$ & $0.4369$ & 24.2 \\
    \hline
    $20^{\circ}$ & $0.3787$ & $0.4668$ & $0.4668$ & $0.4674$ & 23.4 \\
    \hline
    $45^{\circ}$ & $0.4936$ & $0.5665$ & $0.5660$ & $0.5661$ & 14.7 \\
    \hline
    $60^{\circ}$ & $0.6132$ & $0.6720$ & $0.6724$ & $0.6727$ & 9.7 \\
    \hline
    $90^{\circ}$ & $1.0000$ & $1.0000$ & $1.0000$ & $1.0000$ & 0.0 \\
    \hline
  \end{tabular}
  \caption{Values of $\alpha$ for two elliptical disk, with various relative orientation angle between the major axis of the ellipses (column 1), obtained from the analytical formula Eq.~\ref{eq.tensorformula} and using Eq.~\ref{eq:avg} (column 2), and from the numerical calculation using TMFCode (column 3). Column 4 shows the percentage numerical error in $\alpha$ corresponding to $3000^2$ pixels.}
  \label{tle:orient}
\end{table}
\section{$W_2^{1,1}$ for excursion sets on the sphere: application to  simulations of CMB data}\label{sec3}
A CMB field value is associated with each point on a 2-dimensional spherical space. Therefore, the TMFCode is not directly applicable to such data. 
A straightforward way to proceed in this case would be to project the excursion set onto an Euclidean plane and then calculate the TMFs. 

\subsection{Stereographic projection of excursion sets}

\begin{figure}
\centering
\resizebox{1.in}{1.in}{\includegraphics{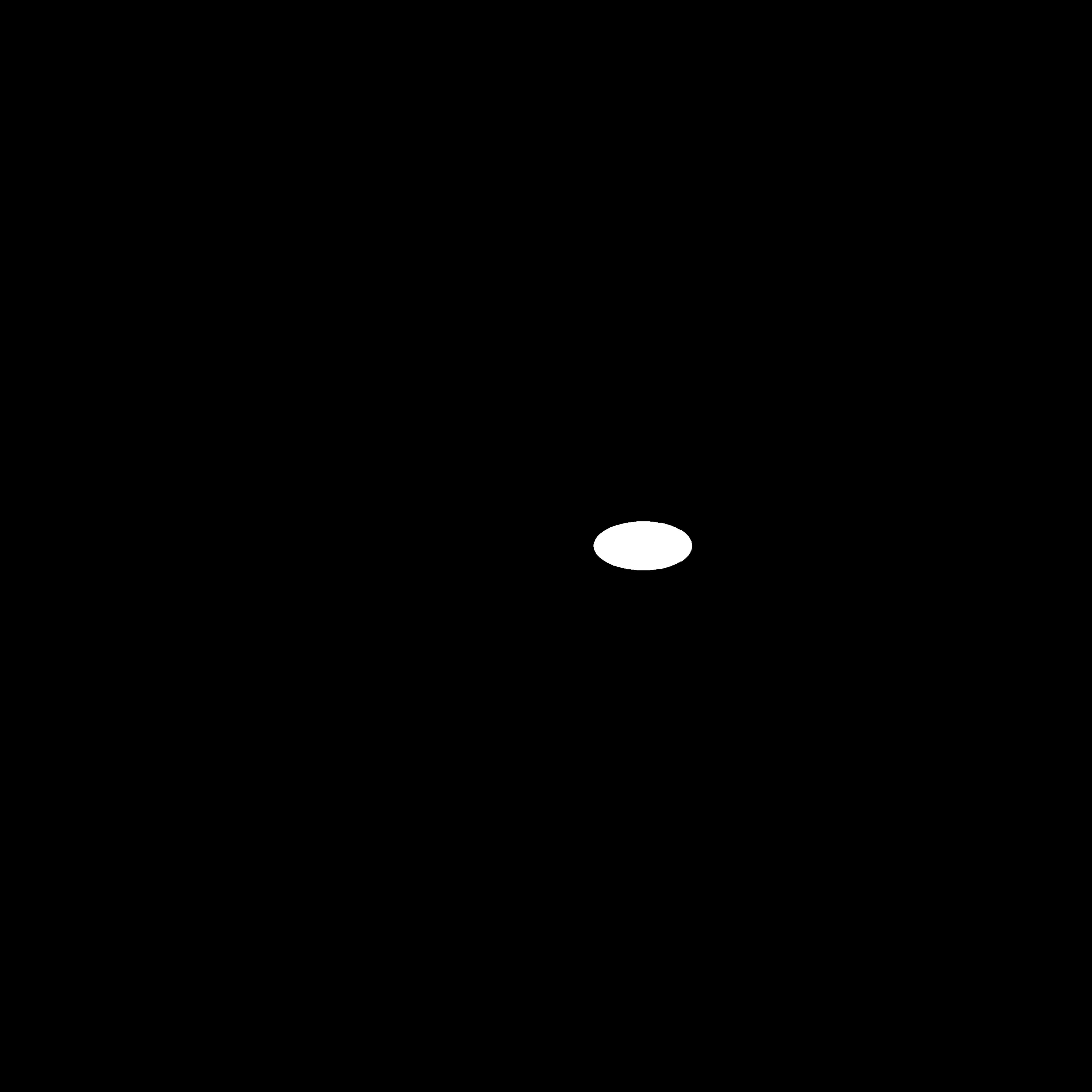}} \quad 
\resizebox{1.in}{1.in}{\includegraphics{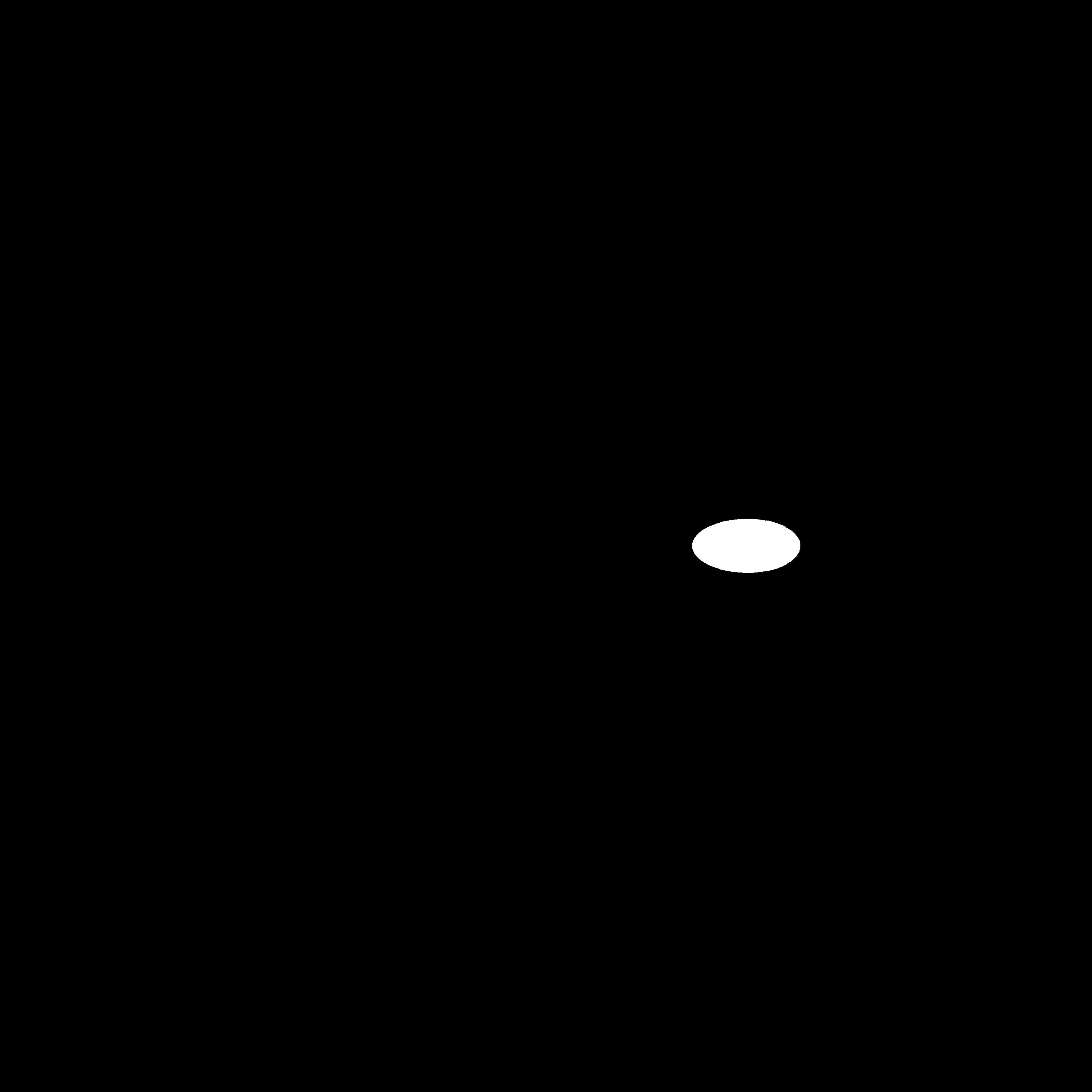}} \quad   
\resizebox{1.in}{1.in}{\includegraphics{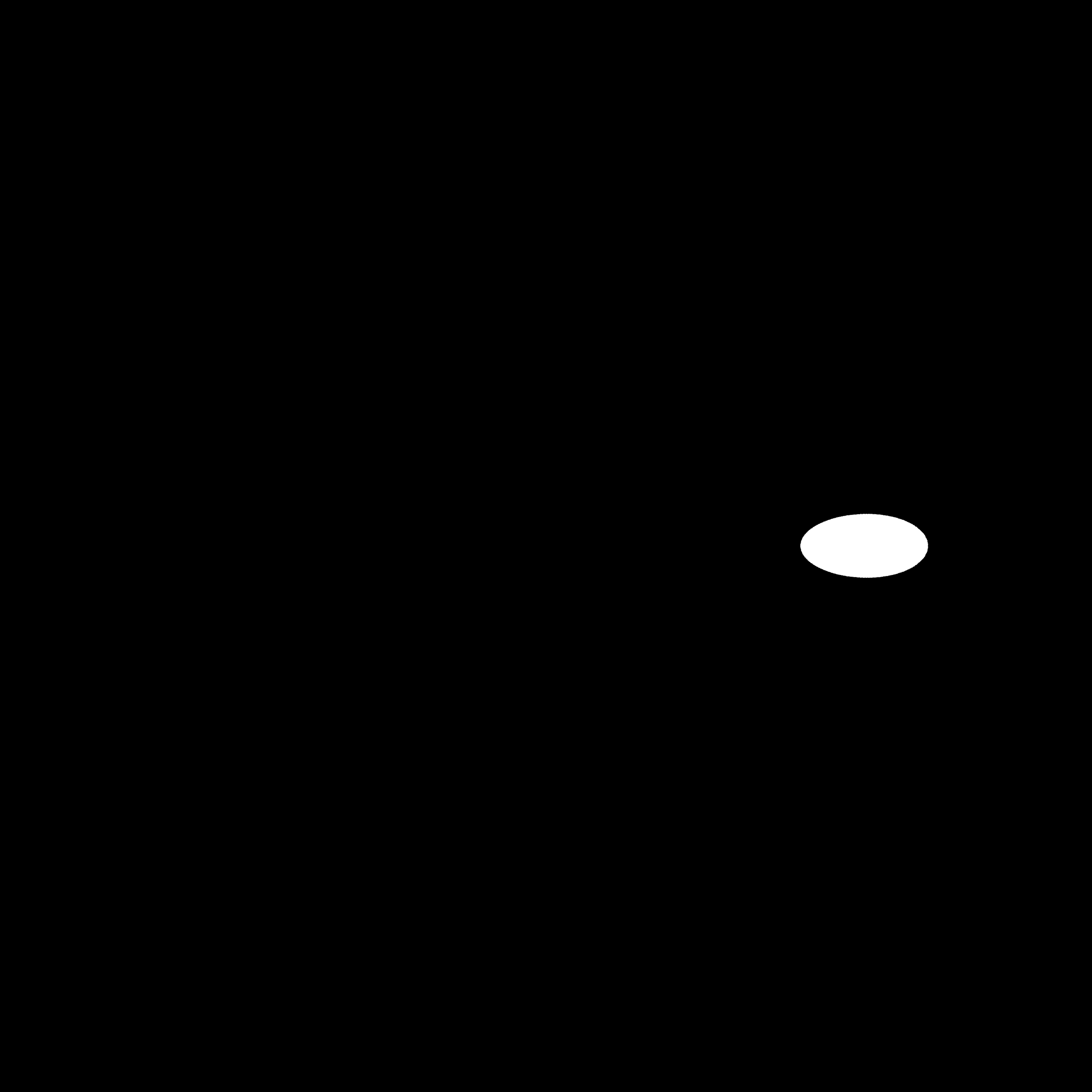}} \quad   
\resizebox{1.in}{1.in}{\includegraphics{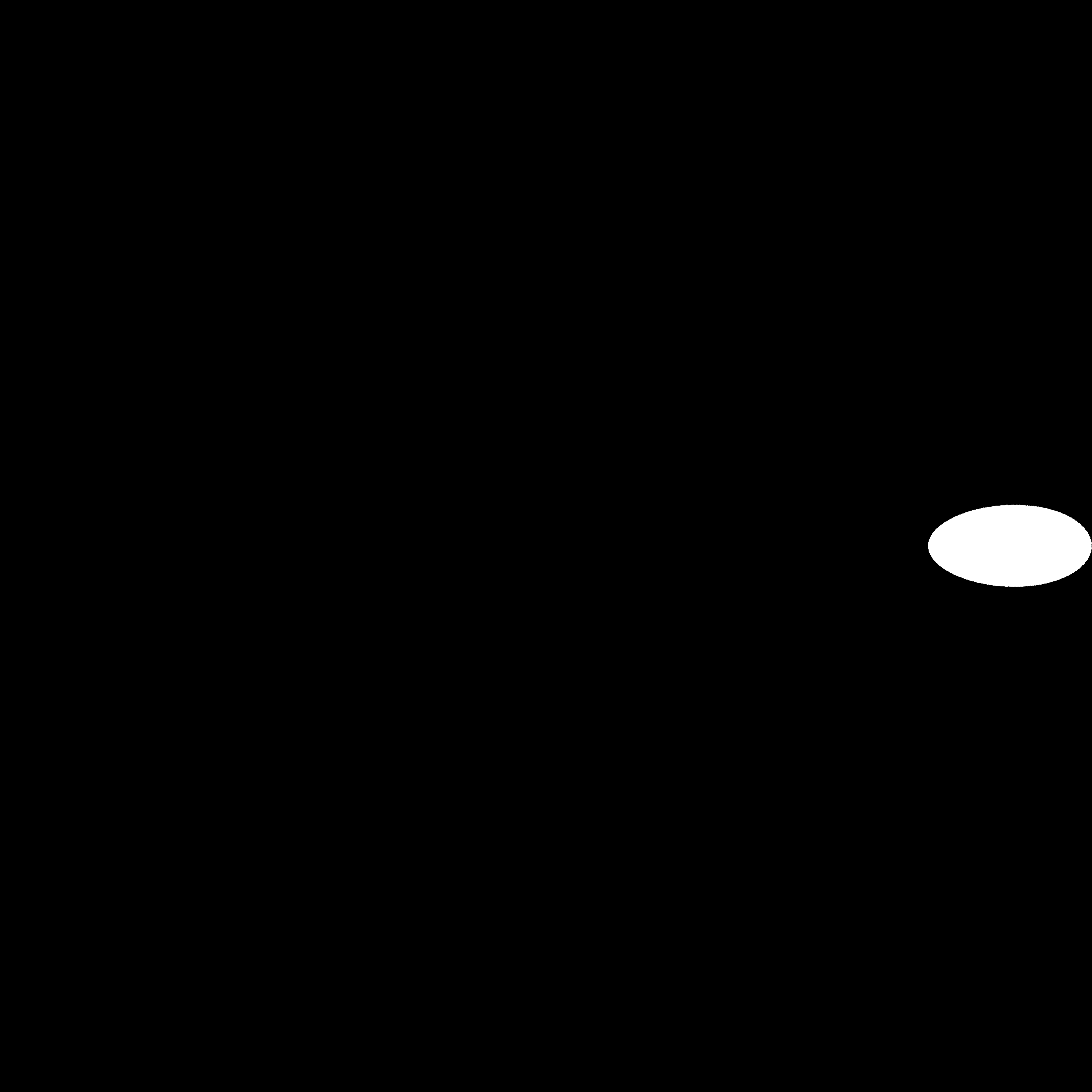}}  
\caption{Stereographic projection image of an elliptic disk (white region), having aspect ratio $q/p$ of $0.5$, located at various positions with fixed azimuthal angle on the sphere. The shape is preserved but the size increases towards the edge of the projection plane.}
\label{fig:deform}
\end{figure}

We choose to use the stereographic projection which is a conformal map and therefore preserves angles and shapes of structures. However, it does not preserve the size. Structures located further away from  the north pole will have projected images on the plane that have the same shape but the size will be scaled up. This can be seen in Fig. \ref{fig:deform} which shows the projection of an elliptic disk with its location on the sphere shifted further and further away from the north pole. Roughly speaking, if $a$ denotes the size of the scaling factor of a structure, then $W_2^{1,1}$ scales as $a$ because $\vec r$ scales as $a$, $\kappa$ as $a^{-1}$, $dl$ as $a$ and the unit normal is invariant. Therefore, two structures that have the same shape but different sizes will have the same value of $\beta$. Hence, stereographic projection is well suited for calculating $\beta$. 

For a collection of structures with a particular distribution of shapes, there are many possible ways in which these structures can be arranged with different relative orientations. In each of these possibilities, the value of $\alpha$ quantifies the extent of orientation in the corresponding collection of structures. This statement remains true for structures on the stereographic projection also, as the shape of the structure remains unaffected as described in the previous paragraph. But due to the stereographic projection, scaling of the structures further away from the north pole leads to errors in the average $\left\langle W_2^{1,1} \right\rangle$ and hence also in the estimation of $\alpha$. However, as the scaling of the structures in the stereographic projection becomes significant as the structures get close to the equator, these errors can be reduced by removing the structures close to the equator.

\vspace{0.5cm}

\subsection{$W_2^{1,1}$ for simulated CMB data}

For a generic random field with zero mean, the set of all pixels having values greater than or equal to a chosen value, usually referred to as {\em threshold}, forms an excursion set. Let the threshold value be denoted by $\nu$. The boundary contours that enclose hotspots and coldspots are iso-threshold contours. The excursion set changes systematically as $\nu$ is varied. Consequently, we can expect the TMFs to follow a corresponding systematic behaviour. 

In this section we calculate $W_2^{1,1}$,  and from it $\alpha$ and $\beta$, for simulated CMB temperature and $E$ mode data, and study the intrinsic shapes and orientations of the structures of the excursion set.  We work with the field rescaled by the corresponding rms value, and so the typical threshold value is of order one. We choose the threshold range $-6<\nu<6$ with 20 bins for our calculations. We focus on Gaussian fields with isotropic distribution of fluctuations. In the absence of analytic expressions for  $\alpha$ and $\beta$ our results here will serve the purpose of providing an estimate of their values that we should expect for Gaussian, isotropic random fields. 

For our analysis we simulate Gaussian and isotropic CMB fields with input $\Lambda$CDM cosmological parameter values given by $\Omega_c h^2$=0.1198, $\Omega_b h^2$=0.02225, $H_0$=67.27, $n_s$=0.9645, $ln(10^{10}A_s)$=3.094 and $\tau$=0.079, from the 2015 Planck data release~\cite{planck:cosmopara2015}. We study the imprint of only the scalar primordial perturbations and hence we simulate $T$ and $E$ fields only. The angular power spectrum for each of the fields is first calculated using the publicly available CAMB package~\cite{Lewis:2000ah}. We take the maximum multipole to be $\ell_{\rm max}=2200$. The $T$ and $E$ maps are then simulated using the HEALPIX~\cite{Gorski:2005} package\footnote{http://healpix.sourceforge.net}.  We choose the HEALPIX resolution parameter value $N_{\rm side}=1024$. We have chosen a Gaussian smoothing of FWHM=$20'$ for $T$, and FWHM=$50'$ for $E$ mode. 

To calculate $W_2^{1,1}$ for each threshold value we apply the procedure outlined in the beginning of this section. The number of pixels on the projected plane is $(3\times N_{\rm side})^2$.  
To minimize numerical inaccuracies arising from structures near the boundary of the stereographic projected disk we remove the structures that fall in the range $\theta$ = $70^{\circ}$ to $110^{\circ}$. To do this we create a map which marks these pixels. Then this is projected on to a plane using stereographic projection. The list of pixels on the projected map which corresponds to the initially marked pixels on the sphere are stored, so that it can be used repeatedly. 
For each $\nu$ structures that overlap with the above pixel list are then removed.
\begin{figure}
  \centering
  \resizebox{4.2in}{3.5in}{\includegraphics{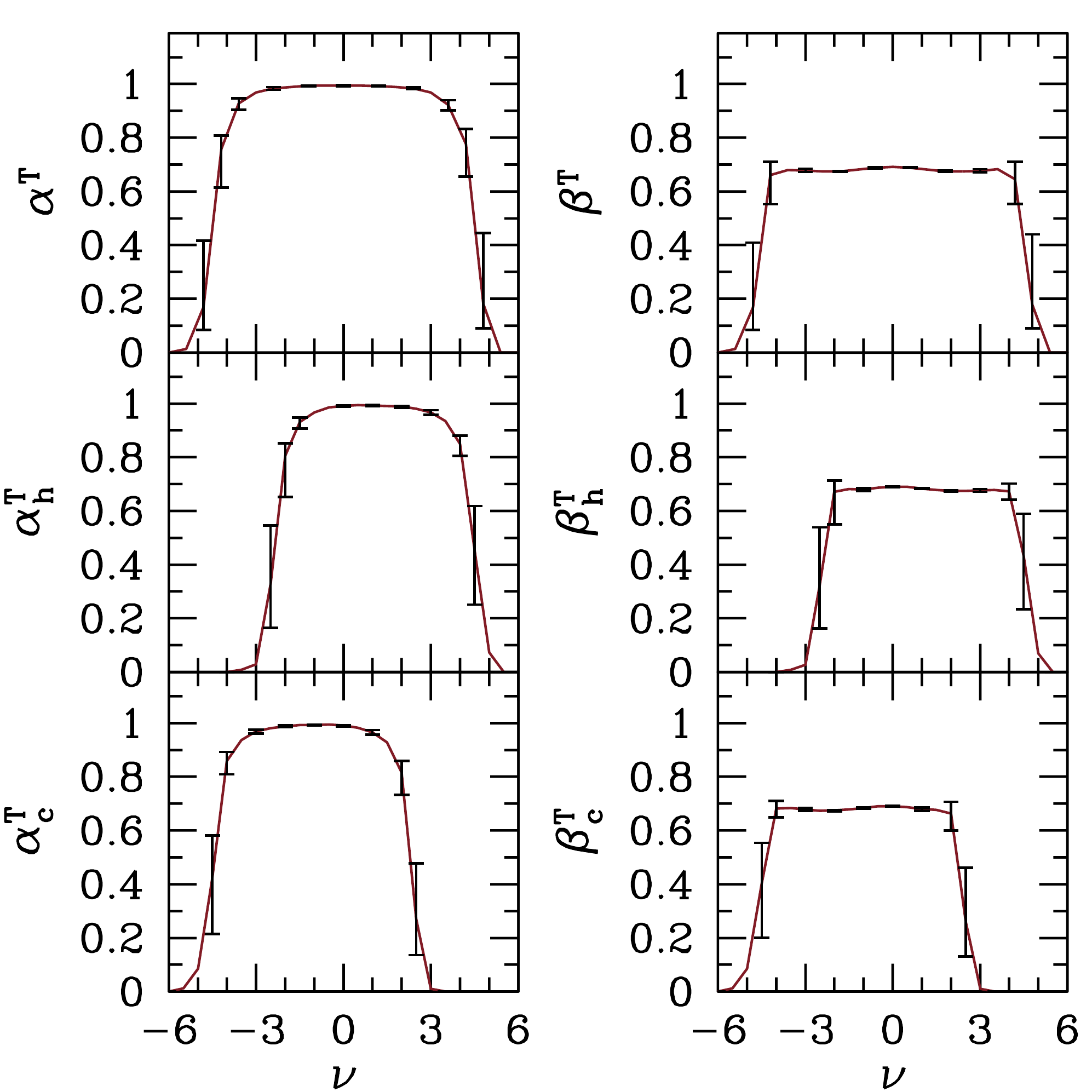}}
  \caption{{\em Top panels:} $\alpha$ and $\beta$ vs $\nu$ calculated using all hotspots and coldspots of the temperature field. {\em Middle panels:} $\alpha$ and $\beta$ vs $\nu$ for hotspots (denoted by subscript $h$) of the temperature field. {\em Bottom panels:} $\alpha$ and $\beta$ vs $\nu$ for coldspots (denoted by subscript $c$) of the temperature field. All plots are average over 100 maps. Error bars are the sample variance from the 100 maps.}
  \label{fig:tnu}
\end{figure}

\begin{figure}
  \centering
  \resizebox{4.2in}{3.5in}{\includegraphics{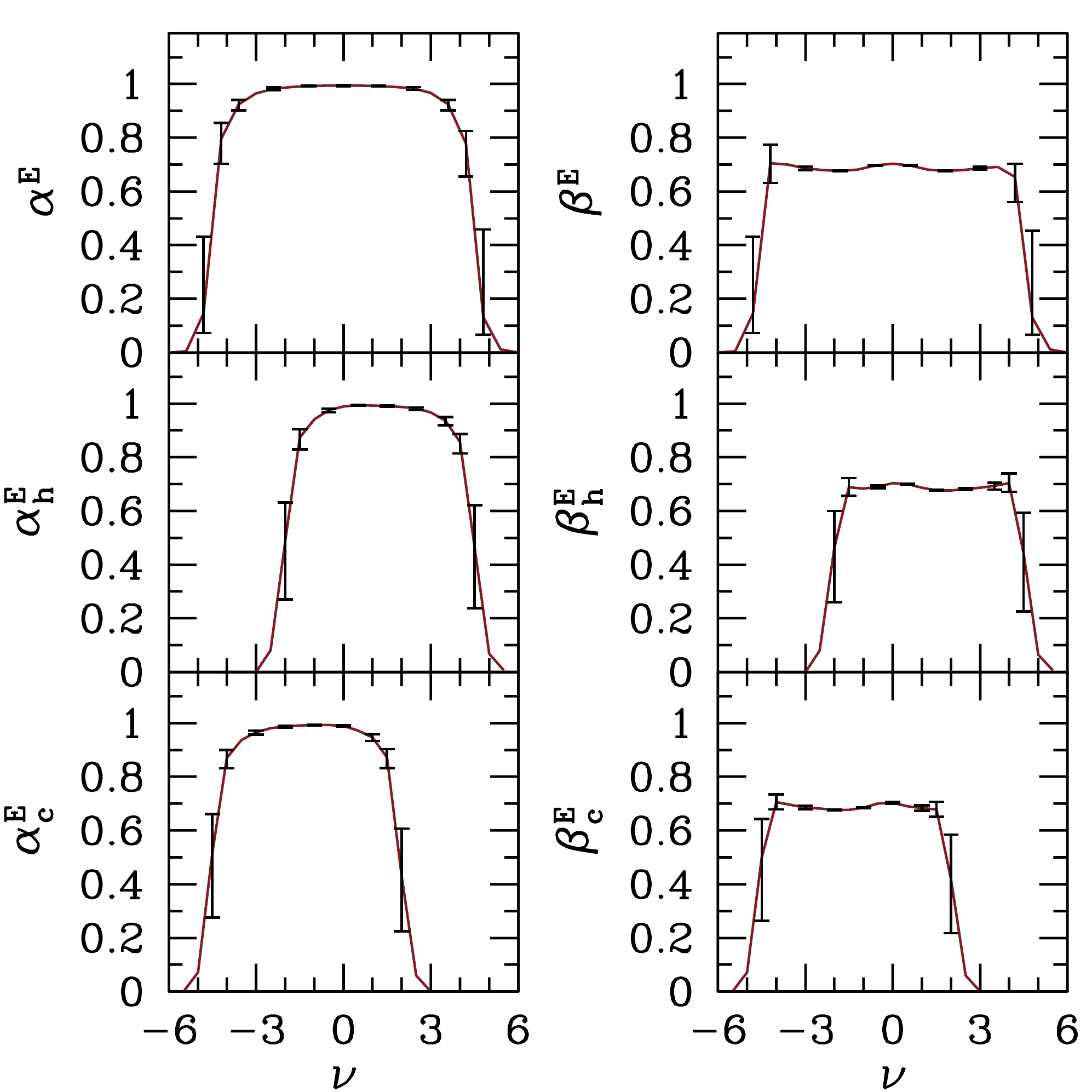}}
  \caption{{\em Top panels:} $\alpha$ and $\beta$ vs $\nu$ calculated using all hotspots and coldspots of the E mode field. {\em Middle panels:} $\alpha$ and $\beta$ vs $\nu$ for hotspots (denoted by subscript $h$) of the E mode field. {\em Bottom panels:} $\alpha$ and $\beta$ vs $\nu$ for coldspots (denoted by subscript $c$) of the E mode field. All plots are average over 100 maps. Error bars are the sample variance from the 100 maps.}
  \label{fig:enu}
\end{figure}

Our results for $\alpha$ and $\beta$ for the temperature field (denoted by superscript $T$) are shown in Fig.~\ref{fig:tnu}. The top left and top right panels show $\alpha^T$  and $\beta^T$ calculated using all hotspots and cold spots. All plots are average over 100 maps. The error bars are the sample variances at each $\nu$ obtained using the 100 realizations. 
We find that $\alpha^T$ is roughly one for $|\nu| < 1$ with small error bars. The conclusion we draw from this is that the simulated CMB temperature maps has {\em no net  orientation} of the structures. This result is consistent with what we expect from a statistically isotropic distribution of a large number of structures, which was our assumption during the map simulation. Note that our result here is obtained for one particular choice of the stereographic projection plane. For a realistic search for net orientation in observed data we will have to work with several choices of projection planes. 
The shape of $\alpha^T$ as a function of $\nu$ appears symmetric about $\nu=0$. $\alpha^T$ becomes less than one for larger values of $|\nu|$ with larger error bars. This can be explained by the fact that for larger $|\nu|$ the number of structures gradually becomes fewer. The reduced sample size of structures leads to a net orientation with the degree of alignment of the structures growing with $|\nu|$. 

From the flat part of the plot of $\beta^T$ we conclude that, the structures are on average {\em intrinsically anisotropic}, and the field has a net anisotropy of roughly $\beta^T \simeq 0.68$. Error bars are small in the flat part of the plot since the number of structures are large, as in the case of $\alpha^T$.  The shape of $\beta^T$ as a function of $\nu$ appears to be symmetric about $\nu=0$. At larger values of $|\nu|$, large number of sample maps do not contain any structures, hence when the $\beta$ from all the sample maps are averaged it results in smaller value of $\beta_{T}$ and larger error bars. The drop from the flat to the decreasing parts are much more sharp in comparison to $\alpha^T$.

We have further calculated $\alpha$ and $\beta$ separately for only hotspots and only coldspots for the temperature field. The results for hotspots (denoted by subscript $h$) are shown in the middle of the left and right panels of Fig.~\ref{fig:tnu}. The corresponding results for coldspots (denoted by subscript $c$) are shown in the bottom part of the left and right panels of the same figure. The `peaks' (central part of the flat region) of the plots for hotspots is located at roughly  $\nu=1$, while for coldspots they are located at roughly $\nu=-1$. These shifts are due to the fact that we find more hotspots for positive values of $\nu$ and more coldspots for negative values. Apart from this shift of the `peaks', the salient features that we observe are similar to what we described above for the combined result of hotspots and coldspots and hence the overall conclusions remain the same.
Hotspots and coldspots independently do not exhibit any net orientation and their net intrinsic anisotropy are the same.

The results for $\alpha$ and $\beta$ for the $E$ mode field (denoted by superscript $E$) are shown in Fig.~\ref{fig:enu}. The behaviour of all the corresponding plots are the same as that was seen earlier for the temperature field. Hence the overall conclusions about the net orientation and intrinsic anisotropy of structures are the same for E mode field. 

\subsection{Correction of the numerical inaccuracy due to pixelization for $\alpha$ and $\beta$ }

The true iso-threshold contours of the excursion sets of the CMB fields are continuous curves, at least up to quantum scales. The results for $\alpha$ and $\beta$ quoted in the previous subsection have been obtained using polygons on pixelised space. As we have shown in Section 2.3.1 the resulting numerical inaccuracy due to the effect of pixelization systematically increases with the rising ellipticity. In order to correct for this inaccuracy we make use of two points. The first is that, upon inspection of a typical CMB temperature or $E$ mode map one can see that contours with very high curvature are rare at any threshold. Secondly, any curved part of a typical contour can be approximated by a portion of an ellipse. Using these two observations we conclude that a reasonable correction of $\beta$ can be made by subtracting the numerical error corresponding to $\beta=0.68$ for a elliptic polygon by interpolating the values in Table~\ref{table:plane}. 
After doing this correction we conclude that the standard $\Lambda$CDM cosmology predicts that the structures of the temperature and $E$ mode CMB fields must have an intrinsic anisotropy given by $\beta=0.62$. This anisotropy encodes the anisotropy of the primordial fluctuations as well as tidal gravitational effects (first order perturbations)  during the recombination epoch. 

As shown in section 2.3.2, the calculation of $\alpha$ has negligible numerical error when the structures are completely unoriented. The value $\alpha=1$ that we obtain here for the simulated CMB fields show that the structures do not have any net orientation and hence no correction of numerical error is needed.  

\subsection{$W_2^{1,1}$ for masked CMB data}

Since the TMFCode is implemented directly on the pixel space it is straightforward to use it for masked CMB fields also. 
The CMB field should not be masked before the projection process as it may result in errors in the estimation of TMFs. If there are structures at the boundary of the mask then masking the CMB field directly without projecting it on a plane, will change the shape of these structures resulting in errors. In order to avoid these errors CMB field and the mask should be separately projected onto the plane. Then the structures in the projected CMB map which overlap with projected masked region can be removed completely without resulting in any errors. The above procedure applies for the case when no smoothing is involved. In the case of the smoothed CMB field, during the smoothing process contaminated pixels at the boundary of mask results in the contamination of the surrounding pixels and the extent to which the contamination spreads depends on the value of the smoothing angle. In order to remove these contaminated pixels, first the mask has to be smoothed with the chosen Gaussian smoothing angle. Then we choose a threshold level whose value represents how far away we stay from the boundary of the original mask map so as to reduce the inaccuracies due to the inclusion of the pixels that is contaminated due to smoothing. Then we change the value of all the pixels which have value above the threshold level into 1 and all the pixels which have value below the threshold level into 0. This will give the final mask for a given smoothing angle. Now the previously described procedure for the case of unsmoothed CMB field can be followed to get the masked projected CMB map for this case also.

\section{Application to CMB data from PLANCK} \label{sec4}

For the analysis in this section we use CMB data from component-separation algorithm SMICA which combines the  observed frequency maps efficiently to minimize the foreground contamination and also the corresponding masks included in the 2015 PLANCK data \cite{PLA}.

In this section we estimate the net orientation and net anisotropy of Planck CMB fields, choosing $\nu=1$ for hotspots and $\nu=-1$ for coldspots. These values of $\nu$ were chosen since the number of structures are the largest around these threshold values. We choose a Gaussian smoothing such that the number of structures are about the order of $10^3$ at the chosen threshold level. This corresponds to a smoothing of FWHM = $20'$ for temperature field and  $50'$ for $E$ mode field. The $1-\sigma$ error bar for the net orientation and net anisotropy can be obtained by averaging over a sample of simulated CMB fields which gives the cosmic variance. 
The simulated CMB maps that we have used for the present analysis does not include any instrumental effects. Therefore, we expect it to have no net orientation and can serve as a reference with which we can compare the value obtained for PLANCK map.

\subsection{Net anisotropy of structures}
Table \ref{ma} shows $\beta$ separately for hotspots and coldspots for PLANCK temperature and $E$ mode data (column 2). 
For comparison, we show $\overline{\beta}$, which is the average over 100 simulation maps (described in Section 3), and the corresponding $1-\sigma$ error bars, in column 4. We find $\beta$ for both temperature and $E$ mode fields of PLANCK to be roughly $0.68$. By applying the correction for pixelization error as was done for the value of $\beta$ calculated from simulated maps in Section 3.2, to temperature and $E$ mode fields for both PLANCK and theoretical expectations from simulations in Table \ref{ma}, we obtain $\beta=0.62$. 

In order to quantify the level of deviations of the measurements for PLANCK data from the theoretical expectations, we define the quantity $\mathscr{D}$ as
\begin{equation}
\mathscr{D} = \left| \frac{X - \overline{X}}{\sigma_X} \right|,
\label{eqn:D}
\end{equation}
where $X$ can be either $\alpha$ or $\beta$ and $\sigma_{X}$ is the corresponding one sigma errorbar.The values of $\mathscr{D}$ for $\beta$ are shown in table \ref{asig}. The calculations shown in tables \ref{ma} and \ref{asig} where obtained for a particular choice of stereographic projection planes. In order to remove any dependence of the results on the choice of the projection plane, we have also repeated the calculations for various choices of projection planes. We find that the value of $\beta$ agrees with $\overline{\beta}$ obtained by averaging over 100 simulation maps to within $3-\sigma$ for all of these various choices of projection planes. In conclusion the net anisotropy of structures of the CMB fields of PLANCK agrees with the theoretical expectations of standard model of cosmology to within $3 \sigma$.

\subsection{Net orientation of structures}

\begin{table}
  \centering
  \begin{tabular}{|l|l|l|p{1.8cm}|l|l|p{1.8cm}|}
    \hline
    Field & \multicolumn{2}{c|}{Planck data} & $\mathscr{O}$ using $\beta$ & \multicolumn{2}{c|}{Average from} & $\mathscr{O}$ using $\beta$ \\
    and & \multicolumn{2}{c|}{} & corrected for & \multicolumn{2}{c|}{100 realizations of} & corrected for \\
    structure & \multicolumn{2}{c|}{} & pixelization & \multicolumn{2}{c|}{simulated data} & pixelization \\
    & \multicolumn{2}{c|}{} & error & \multicolumn{2}{c|}{with $1-\sigma$ error bars} & error \\
    \cline{2-3}\cline{5-6}
    &&&&&& \\
    & $\alpha$ & $\beta$ && $\overline{\alpha}$ & $\overline{\beta}$ & \\
    \hline
    Temperature & $0.9889$ & $0.6795$ & $0.0290$ & $0.9911^{+0.0034}_{-0.0054}$ & $0.6754^{+0.0026}_{-0.0030}$ & $0.0231$ \\
    hotspot &&&&&& \\
    \hline
    Temperature & $0.9936$ & $0.6791$ & $0.0167$ & $0.9910^{+0.0038}_{-0.0052}$ & $0.6754^{+0.0030}_{-0.0026}$ & $0.0233$ \\
    coldspot &&&&&& \\
    \hline
    E mode & $0.9673$ & $0.6820$ & $0.0861$ & $0.9930^{+0.0034}_{-0.0034}$ & $0.6858^{+0.0022}_{-0.0028}$ & $0.0186$ \\
    hotspot &&&&&& \\
    \hline 
    E mode & $0.9593$ & $0.6812$ & $0.1069$ & $0.9928^{+0.0028}_{-0.0038}$ & $0.6854^{+0.0032}_{-0.0030}$ & $0.0191$ \\
    coldspot &&&&&& \\
    \hline
  \end{tabular}
  \caption{{\em Column 2:} $\alpha, \beta$ at a chosen threshold, for temperature and $E$ mode data from PLANCK calculated separately for hotspots and coldspots. The threshold value chosen was $\nu=1$ for hotspots and $\nu=-1$ for coldspots. {\em Column 3 :} $\mathscr{O}$ defined in Eq.~\ref{eqn:O} calculated using the values of $\alpha$, and $\beta$ corrected for pixelization error. {\em Column 4:} The corresponding average values (denoted by over bar) calculated from 100 simulated temperature and $E$ mode maps. The corresponding $1 \sigma$ error bars calculated using 100 simulations. {\em Column 5:} The corresponding value of $\mathscr{O}$ calculated using average values $\overline{\alpha}$ and $\overline{\beta}$ corrected for pixelization error.}
  \label{ma}
\end{table}

Table \ref{ma} shows $\alpha$ separately for hotspots and coldspots for PLANCK temperature and $E$ mode data (column 2). 
For comparison, in column 4 we show $\overline\alpha$, which is the average over 100 simulation maps (described in Section 3), and the corresponding $1\sigma$ error bars. Comparison of the level of orientation of the structures in various fields can be conveniently done using $\mathscr{O}$ defined in Eq.~\ref{eqn:O}. $\mathscr{O}$ calculated using $\alpha$ and with $\beta$ corrected for pixelization effect, from PLANCK data are shown in column 3 of Table.~\ref{ma}. The calculations of $\mathscr{O}$ corresponding to $\overline{\alpha}$ and with $\overline{\beta}$ corrected for pixelization effect, which are the averages over $100$ simulation maps are shown in column 5. We find that the extent of orientation in the structures of the $E$ mode field, $\mathscr{O}$, from PLANCK data is large compared to what is obtained from the average over many realizations while the corresponding values obtained for temperature field are comparable. 
Table~\ref{asig} shows $\mathscr{D}$ calculated for hotspots and coldspots for temperature and $E$ mode fields.  It can be observed that for temperature field, all values of $\mathscr{D}$ are less than one. However, for $E$ mode the values are much larger than one. 
For these calculations we had chosen the Galactic plane as the stereographic projection plane.

The results described above were obtained for a particular choice of projection plane. We then ask whether these results are a fluke arising due to the choice of stereographic projection plane (described in \ref{A1}). So in order to further analyze the net orientation for temperature and $E$ mode fields, we have repeated the calculations for several other choices of projection planes. We still find that the deviation of $\alpha$ from $\overline{\alpha}$ for the temperature field is within $3-\sigma$, but for $E$ mode the deviation is significant and of about $14-\sigma$ from the theoretical expectation of no net orientation. Hence, we conclude that the structures in the $E$ mode data from PLANCK do exhibit a net orientation. 
The reason behind this behaviour may be due to the instrumental effects that has not been included in the simulated CMB maps or this may be due to the contamination present in the PLANCK $E$ mode map. Or this may be a signature of the existence of some net orientation in the structures of the $E$ mode field. The exact reason can only be revealed by further investigation.  

\begin{table}
  \centering
  \begin{tabular}{|p{3.5cm}|p{3.1cm}|p{3.1cm}|}
    \hline
    && \\
    Field and structure & $\mathscr{D} = \left| \dfrac{\alpha - \overline{\alpha}}{\sigma_{\alpha}} \right|$ & $\mathscr{D} = \left| \dfrac{\beta - \overline{\beta}}{\sigma_{\beta}} \right|$ \\
    && \\
    \hline
    Temperature & $0.5000$ & $1.4643$ \\
    hotspots && \\
    \hline
    Temperature & $0.5778$ & $1.3214$ \\
    coldspots && \\
    \hline
    $E$ mode & $7.5588$ & $1.5200$ \\
    hotspots && \\
    \hline
    $E$ mode & $10.1515$ & $1.3548$ \\
    coldspots && \\
    \hline
  \end{tabular}
  \caption{The table shows the quantity $\mathscr{D}$, defined in Eq.~\ref{eqn:D}, for $\alpha$ (column 2) and $\beta$ (column 3).}
  \label{asig}
\end{table}

\section{Conclusion} \label{sec5}
We have introduced tensor Minskowski Functionals as a new statistic to analyze CMB data. The new information about the morphology of structures, namely, the net orientation and net anisotropy of structures, opens up the possibility for a wide range of applications in cosmology. The applications include new tests for the standard cosmological model and searches for deviations from it,  
resolving the anomalies in the CMB data, and to constrain the physics of the early Universe. They can also be used to understand characteristic signatures of instrumental and foreground signals. They are also promising  for the analysis of data of the large scale structure  and 21 cm emissions from the epoch of reionization. 

We have developed a code for calculating TMFs, focussing on the $(1,1)$  rank tensor $W_2^{1,1}$ which is a generalization of the genus, for excursion sets of two dimensional random fields. The code then uses  $W_2^{1,1}$ to measure the net orientation, $\alpha$, and net intrinsic anisotropy, $\beta$, of structures. Our intended application is to random fields on continuous space and the pixelization of the data introduces numerical errors. We have done a careful estimation of this numerical error by using the known formula for $W_2^{1,1}$ for  ellipses. We find that the error in the measure of the intrinsic anisotropy increases as the curvature of the boundaries of the structure increases. For the net orientation we find that for a distribution of structures that are completely unaligned with each other, the error is negligible. However, as the structures become more and more aligned, the error approaches the value of the error for the intrinsic anisotropy. 

CMB data is associated with each point on the spherical sky. So in order to apply the TMFCode, we use stereographic projection of the CMB field onto a plane. We calculate $W_2^{1,1}$, and then compute $\alpha$ and $\beta$ as functions of different threshold levels for simulated Gaussian and isotropic CMB temperature and $E$ mode fields. We find that the standard $\Lambda$CDM predicts that the level of intrinsic anisotropy of hotspots and coldspots in both the CMB fields to be $\beta = 0.62$, where correction due to pixelization has been taken into account. Further, we find the value of $\alpha$ to be one for botih temperature and $E$ mode fields, which implies that there is no net orientation in the structures of these fields. This is a recovery of the statistical isotropy of density fluctuations that we have input into the CMB simulations.
Then, we use TMFCode to compute $\alpha$ and $\beta$ for temperature and $E$ mode data from PLANCK mission. We find that $\beta$ for both temperature and $E$ mode data are consistent with the expectations from standard $\Lambda$CDM simulations within $3-\sigma$. Further, we find that the temperature field agrees with the standard $\Lambda$CDM prediction of no net orientation within $3-\sigma$. However, we find $14-\sigma$ evidence for a net orientation in $E$ mode data. The reason behind this may be instrumental effects that we have not included in the simulated CMB maps, or this may be due to the contamination present in the PLANCK $E$ mode data. This may also be a signature of the existence of some net orientation in the structures of the $E$ mode field. The exact reason can only be revealed by further investigation and we plan to come back to this issue after the PLANCK team releases the full polarization data. 

This work is the initiation of several lines of investigation. We list here some of our ongoing and planned work. We have ignored tensor perturbations, and consequently $B$ mode field in this work. We are currently studying the net orientation and net anisotropy for $B$ mode sourced by primordial tensor perturbation as well as sourced by lensing due to the large scale structure of the Universe. Further, non-Gaussianity of primordial fluctuations generated during the very early Universe has an imprint on the CMB fields. Understanding of the effect of primoridial non-Gaussianity on the TMFs of CMB fields may eventually lead to an alternative tool to constrain $f_{NL}$ which can be used as a consistency check. And with further development it may be used to distinguish different origins of non-Gaussianity in CMB fields. We are currently pursuing this issue. 

The effects which are of primordial origin should be dominant on angular scales larger than roughly $1^{\circ}$. We plan to use this fact to study whether TMFs of CMB fields are capable of distinguishing the primordial origin of fluctuations from the late time effects. 
Further, our study here was carried out for one set of cosmological parameters. It would be important to understand the effect of various cosmological parameters on the TMFs of the CMB fields. 
This will lead to a better understanding of the power of using TMFs as a tool to constrain cosmology. Since a precise measurement of $\alpha$ and $\beta$ is necessary for model testing and constraining, it is very important to get a better handle on the numerical inaccuracies induced due to the pixelization and we are searching for ways to minimize them. 

Since the TMFs are quite general and applicable to any random field on dimensions higher than or equal to two, they can be applied to other areas of cosmology as well. The two-dimensional case developed here can be used to study 21cm emissions from the epoch of reionization on two-dimensional redshift slices in a straightforward manner. This can be used to probe cosmological parameters and also different models of reionization. 
Further, we plan to develop codes which can calculate TMFs for 3-dimensional data~\cite{schroder3D:2013} with the aim of constraining cosmology using $21$cm data from future radio interferometers and large scale structure data.

\ack{We acknowledge the use of the Hydra cluster at the Indian Institute of Astrophysics. Some of the results in this paper have been obtained by using the \texttt{CAMB}~\cite{Lewis:2000ah,cambsite} software and HEALPIX~\cite{Gorski:2005,Healpix} package for generating the simulated CMB maps. P.C would like to thank K.~P.~Yogendran for useful discussions.} 
\appendix
\section{Orientation measure of the structures of excursion set on a sphere}
\label{A1}
The orientation measure $\alpha$ can detect the existence of any alignment in structures of the excursion set on a plane (described in Section 2.3.2). In this section we extend this study to the excursion set on a sphere. We consider an excursion set on the sphere with many ellipses whose aspect ratio are fixed to be 0.7. We consider all of these ellipses to be aligned towards the pole of the sphere and hence has net orientation. In order to reduce the numerical inaccuracies due to the stereographic projection we remove the ellipses which fall in the range $\theta = 70^{\circ}$ to $110^{\circ}$.

\begin{figure}
\centering
\resizebox{1.5in}{1.5in}{\includegraphics{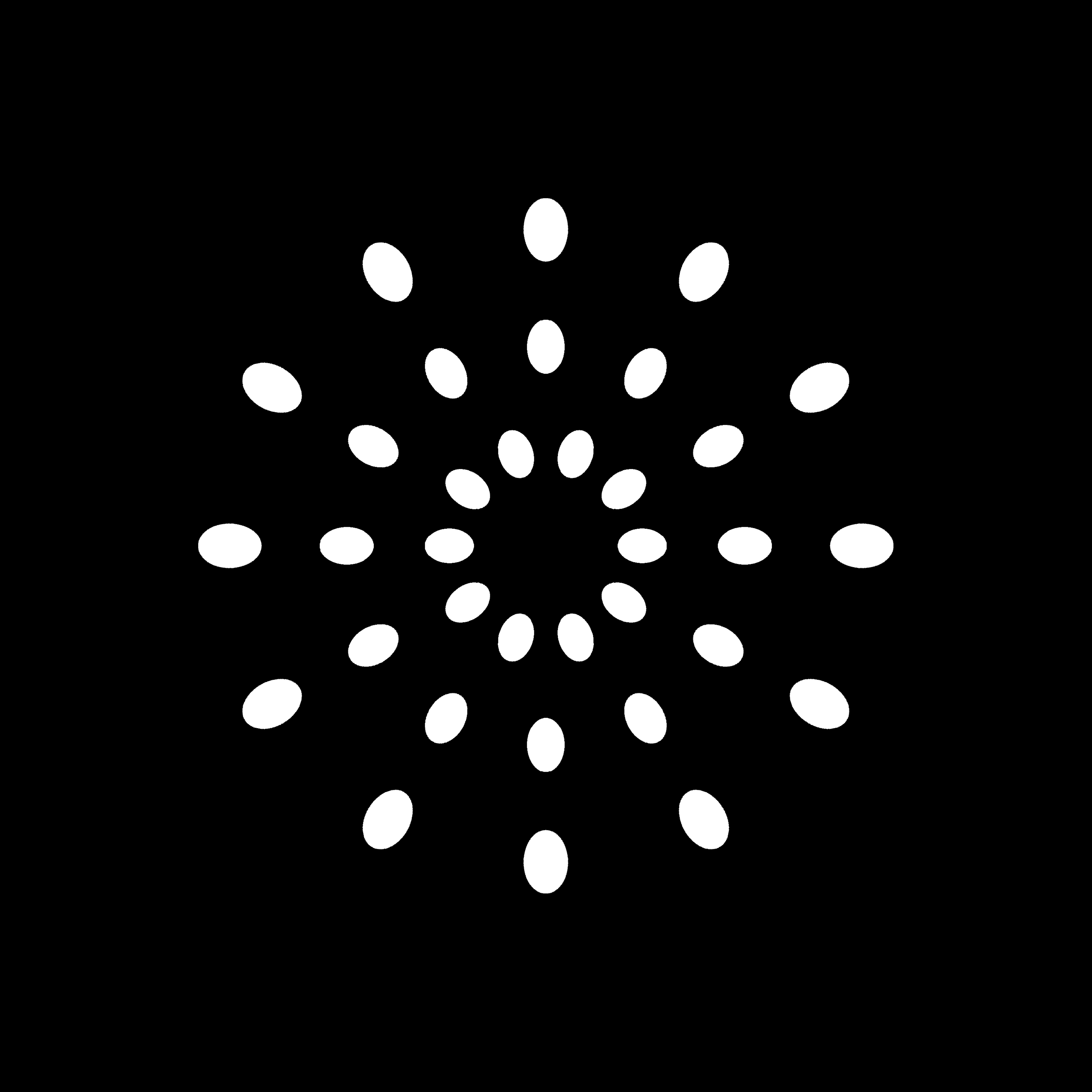}} \quad \quad \resizebox{1.5in}{1.5in}{\includegraphics{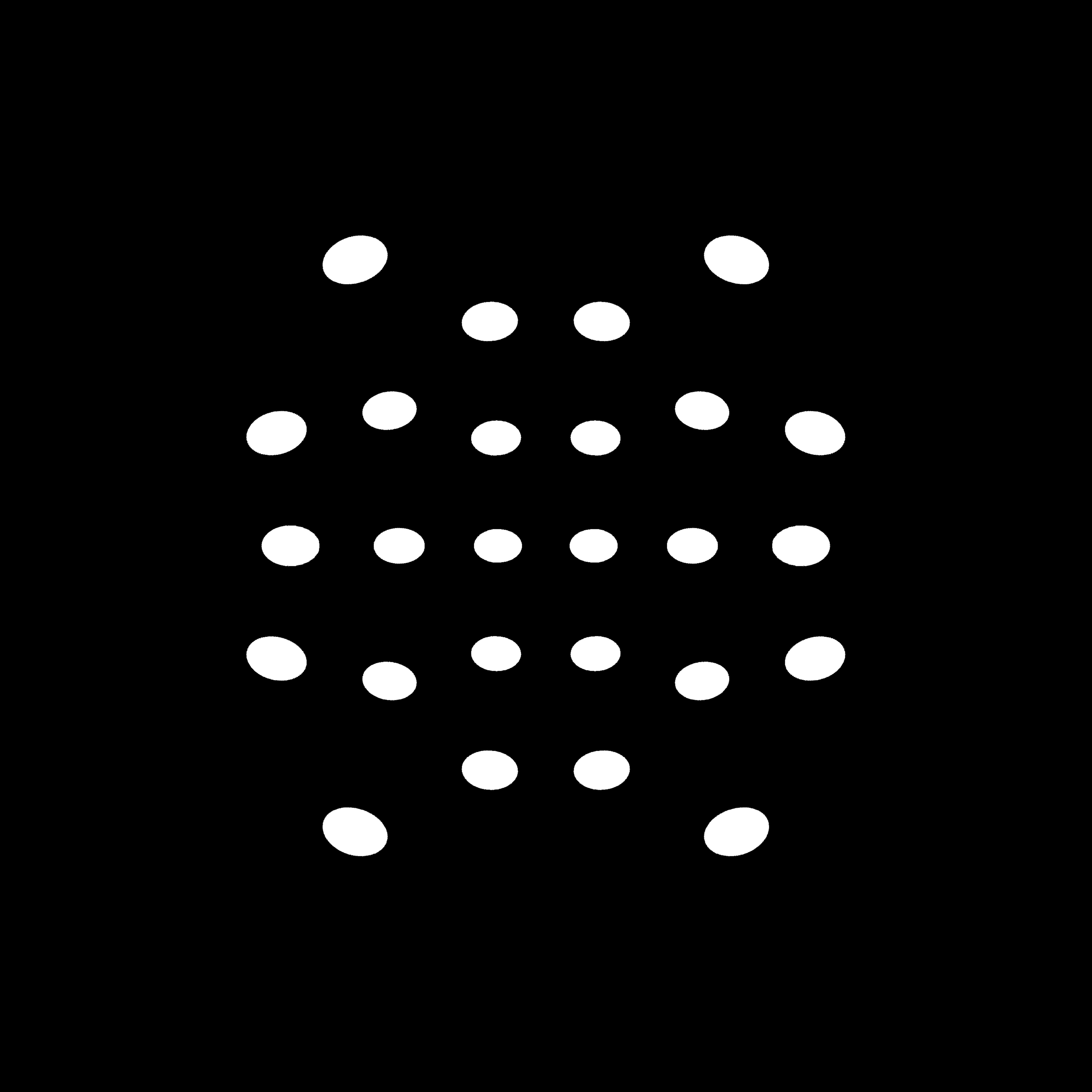}} \\
\caption{Stereographic projection image of the excursion set on a sphere with many ellipses. The image on the left shows the stereographic projection for the case when the equator of unit sphere with the north pole $(0^{\circ},0^{\circ})$ is the projection plane. And the image on the right corresponds to the case where $(90^{\circ},0^{\circ})$ is the north pole.}
\label{fig:orient}
\end{figure}

Table \ref{table:ellipse_orient} shows the net orientation $\alpha$ and the normalized form of net orientation, $\mathscr{O}$ defined in Eq.~\ref{eqn:O}, for various choices of stereographic projection plane. In each of these choices the equator of the unit sphere with the north pole specified in column 1 of Table \ref{table:ellipse_orient} is the projection plane. We find that $\alpha$ and $\mathscr{O}$ does not vary significantly with $\varphi$ but varies with $\theta$. At $\theta = 0^{\circ}$, the value of $\alpha$ and $\mathscr{O}$ shows that the structures on the corresponding projection plane has no net orientation. While for the cases with $\theta=90^{\circ}$, the values show that the structures on the corresponding projection plane has a high level of orientation. The stereographic projection image corresponding to these two cases are shown in Fig.~\ref{fig:orient}. We find that these images also show the same results, that the structures corresponding to $\theta=0^{\circ}$ are not oriented with each other while for the case of $\theta=90^{\circ}$ shows a net orientation. Hence we can come to the conclusion that even though the structures of the excursion set on the sphere has net orientation, its stereographic projection onto a particular choice of projection plane may or may not show a net orientation. While for the case where the structures of the excursion set are randomly oriented then for any choice of the projection plane, the structures on these plane will also be randomly oriented and hence will have no net orientation. So in order to detect the orientation in the structures of the excursion set on a sphere one has to analyze the orientation for various choices of stereographic projection planes. Further we can infer the extent of orientation in the structures of the excursion set on the sphere by estimating $\alpha$ corresponding to the projection plane which shows maximum net orientation. In the present case of excursion set on a sphere with many ellipses, this corresponds to the choice of projection plane with $\theta = 90^{\circ}$ and, hence $\alpha = 0.74, \mathscr{O} = 0.93$. The value of $\mathscr{O}$ is close to one which implies that the structures in the present case of the excursion set are highly oriented with each other which is the result that we expected. Finally in conclusion, the orientation in the structures of the excursion set on the sphere can be detected and its extent of orientation can be quatified by estimating $\alpha$ for various choices of stereographic projection planes.

\begin{table}
  \vspace{0.5cm}
  \centering{}
  \begin{tabular}{| p{3.5cm} | p{2.5cm} | p{2.5cm} |}
    \hline
    North pole  & $\alpha$ & $\mathscr{O}$ \\   
     $(\theta,\varphi)$ && \\
    &&\\
    \hline 
    ($0^{\circ}$,$0^{\circ}$) & $0.9963$ & $0.0107$ \\
    \hline
    ($45^{\circ}$,$0^{\circ}$) & $0.9322$ & $0.2204$ \\
    \hline
    ($45^{\circ}$,$90^{\circ}$) & $0.9212$ & $0.2529$ \\
    \hline
    ($45^{\circ}$,$180^{\circ}$) & $0.9322$ & $0.2204$ \\
    \hline
    ($45^{\circ}$,$270^{\circ}$) & $0.9212$ & $0.2529$ \\
    \hline
    ($90^{\circ}$,$0^{\circ}$) & $0.7259$ & $0.9259$ \\
    \hline
    ($90^{\circ}$,$45^{\circ}$) & $0.7368$ & $0.9236$ \\
    \hline
    ($90^{\circ}$,$90^{\circ}$) & $0.7208$ & $0.9310$ \\
    \hline
    ($90^{\circ}$,$135^{\circ}$) & $0.7368$ & $0.9236$ \\
    \hline
  \end{tabular}
  \caption{The table showing the net orientation $\alpha$ (column 2) and $\mathscr{O}$ (column 3) defined in Eq.~\ref{eqn:O}, for various choices of projection plane, for the excursion set on a sphere with many ellipses. In each of these various choices the equator of the unit sphere with the north pole specified in column 1 is the chosen projection plane.}
  \label{table:ellipse_orient}
\end{table}

\end{document}